\documentclass[draftclsnofoot,12pt,onecolumn]{IEEEtran}   
\usepackage{setspace}
  \usepackage[dvips]{graphicx}
  \DeclareGraphicsExtensions{.eps}
\usepackage[cmex10]{amsmath}
\usepackage{gensymb}

\usepackage{array,booktabs}%

\usepackage{courier}

\begin{document}
%
\title{Second-Order Statistics of \emph{$\kappa-\mu$} Shadowed Fading Channels}
%
%
%

\author{Simon L. Cotton,~\IEEEmembership{Senior Member,~IEEE}
\thanks{This work was supported by the U.K. Royal Academy of Engineering, the Engineering and Physical Sciences Research Council (EPSRC) under Grant References EP/H044191/1 and EP/L026074/1, and also by the Leverhulme Trust, UK through PLP-2011-061.}
\thanks{S. L. Cotton is with the Institute of Electronics, 
Communications and Information Technology, 
Queen's University Belfast, UK, BT3 9DT 
(phone: +44 2890971749; fax: +44 2890971702; e-mail: simon.cotton@qub.ac.uk).}
\thanks{This work has been submitted to the IEEE for possible publication. Copyright may be transferred without notice, after
which this version may no longer be accessible.}}

\maketitle

\begin{abstract}
\normalsize{
\begin{spacing}{1.5}
\noindent In this paper, novel closed-form expressions for the level crossing rate (LCR) and average fade duration (AFD) of $\kappa-\mu$ shadowed fading channels are derived. The new equations provide the capability of modeling the correlation between the time derivative of the shadowed dominant and multipath components of the $\kappa-\mu$ shadowed fading envelope. Verification of the new equations is performed by reduction to a number of known special cases. It is shown that as the shadowing of the resultant dominant component decreases, the signal crosses lower threshold levels at a reduced rate. Furthermore, the impact of increasing correlation between the slope of the shadowed dominant and multipath components similarly acts to reduce crossings at lower signal levels. The new expressions for the second-order statistics are also compared with field measurements obtained for cellular device-to-device and body centric communications channels which are known to be susceptible to shadowed fading.
\end{spacing}
}  
\end{abstract}


\pagebreak

%
\IEEEpeerreviewmaketitle

\section{Introduction}
%
%
%
%
\noindent The $\kappa-\mu$ shadowed fading model first appeared in the literature in \cite{1} and immediately after this in \cite{2}. It has been proposed as a generalization of the popular $\kappa-\mu$ fading model \cite{3}. In this model clusters of multipath waves are assumed to have scattered waves with identical powers, alongside the presence of elective dominant signal components -- a scenario which is identical to that observed in $\kappa-\mu$ fading \cite{3}. The key difference between the $\kappa-\mu$ shadowed fading model and that of classical $\kappa-\mu$ fading is that the dominant components of all the clusters can randomly fluctuate because of shadowing. In particular it is assumed that the shadowing fluctuation follows a Nakagami distribution \cite{4}. Like the $\kappa-\mu$ distribution, the $\kappa-\mu$ shadowed distribution is an extremely versatile fading model which also contains as special cases other important distributions such as the One-Sided Gaussian, Rice (Nakagami-$n$), Nakagami-$m$ and Rayleigh distributions. In addition to this, it also contains as a special case Abdi's signal reception model \cite{5} which considers Ricean fading where the dominant component also undergoes shadowed fading which follows the Nakagami distribution. Due to the ability of the Nakagami probability density function (PDF) to approximate the lognormal PDF \cite{6}, the $\kappa-\mu$ shadowed fading model can also be used to estimate Loo's well-known model for land mobile satellite communications \cite{7}.

While the research of composite fading models such as the $\kappa-\mu$ / gamma model \cite{8} and its associated second-order statistics including the level crossing rate (LCR) and average fade duration (AFD) \cite{9} have been advanced, unfortunately, at present, similar closed-form expressions for the second-order statistics of the $\kappa-\mu$ shadowed model are currently unavailable in the literature. The LCR and AFD of a fading signal are of great importance in the design of mobile radio systems and in the analysis of their performance \cite{10}. Among their many potential applications are the design of error correcting codes, optimization of interleaver size and system throughput analysis as well as channel modeling and simulation. In this paper, convenient closed-form expressions for the level crossing rate and average fade duration of $\kappa-\mu$ shadowed fading channels are derived and subsequently verified by reduction to known special cases. An important empirical validation is also performed through comparison with field measurements from two different types of wireless channel which are known to suffer from shadowed fading, namely cellular device-to-device (D2D) communications channels \cite{2} and body centric communications channels \cite{11}.

The remainder of this paper is organized as follows. Section II provides a brief overview of the $\kappa-\mu$ shadowed fading model. Important relationships between the $\kappa-\mu$ shadowed fading envelope and its time derivative, which underlie the second-order equations proposed here, are established in Section III. Also presented in Section III is the derivation of the LCR, while the derivation of the AFD is given in Section IV. The new expressions for the LCR and AFD of the $\kappa-\mu$ shadowed fading model are compared with some empirical data obtained from field measurements in Section V. Lastly, Section VI finishes the paper with some concluding remarks.

\section{An Overview of the \emph{$\kappa-\mu$} Shadowed Fading Model}
%
%
%
%
\noindent The $\kappa-\mu$ shadowed fading model was originally proposed in \cite{1}. A slight variant of the underlying signal model was also developed independently and appeared shortly after this in \cite{2}. In \cite{1} a rigorous mathematical development of the model was performed, while in \cite{2} the model was the result of channel measurements conducted to characterize the shadowed fading observed in device-to-device communications channels. Both papers have developed important statistics related to the $\kappa-\mu$ shadowed fading model such as the probability density function and moment generating function. In \cite{1}, the cumulative distribution function (CDF) and the sum and maximum distributions of independent but arbitrarily distributed $\kappa-\mu$ shadowed variates were also derived, while, the moments of this model were presented in \cite{2}. In the sequel, the model presented in \cite{1} is used to develop the LCR and AFD equations proposed here. It is worth highlighting that the models proposed in \cite{1} and \cite{2} are related by a simple scaling factor applied to the dominant signal component and thus either could be used to arrive at the second-order equations presented here.

The received signal envelope, $R$, of the $\kappa-\mu$ shadowed fading model may be expressed in terms of the in-phase and quadrature components of the fading signal such that \cite{1}
 
\begin{equation}
R^{2} = \sum^{\mu}_{i=1} (X_{i} + \xi p_{i})^2 + (Y_{i} + \xi q_{i})^2
\end{equation}

\noindent where $\mu$ is the number of multipath clusters, which is initially assumed to be a natural number\footnote{Note, this restriction is later relaxed by allowing $\mu$ to assume any positive real value.}, $X_{i}$ and $Y_{i}$ are mutually independent Gaussian random processes with mean $E[X_{i}] = E[Y_{i}] = 0$ and variance $E[X^{2}_{i}] = E[Y^{2}_{i}] = \sigma^{2}$ (i.e. the power of the scattered waves in each of the clusters). Here $p_{i}$ and $q_{i}$ are the mean values of the in-phase and quadrature phase components of multipath cluster $i$ and $d^{2} = \sum^{\mu}_{i=1} p^{2}_{i} + q^{2}_{i}$. In this model, all of the dominant components are subject to the same common shadowing fluctuation, $\xi$, which is a Nakagami-$m$ random variable with the shaping parameter $m$ used to control the amount of shadowing experienced by the dominant components and $E[\xi^{2}] = 1$. As with the $\kappa-\mu$ model \cite{3}, $\kappa > 0$ is simply the ratio of the total power of the dominant components ($d^{2}$) to the total power of the scattered waves (2$\mu\sigma^{2}$) and the mean power is given by $E[R^{2}] = \bar{r}^{2} = d^{2} + 2 \mu\sigma ^ {2}$. While the PDF of $\it{R}$, $\it{f}_{R}$($\it{r}$), could be obtained from \cite[eq. (8)]{2} by expressing the mean power of the dominant component ($\Omega$) in terms of $\kappa$ and $\bar{r}$, that is $\Omega = \kappa\bar{r}^{2}/(\kappa+1)$, for the purposes of this derivation it is obtained from the PDF of the instantaneous signal-to-noise ratio ($\gamma$) given in \cite[eq. (4)]{1} via a transformation of variables ($\gamma = {r}^2\bar{\gamma}/\bar{r}^2$) as

\begin{equation}
{f_R}(r) = \frac{{2{r^{2\mu  - 1}}{\mu ^\mu }{m^m}{{\left( {1 + \kappa } \right)}^\mu }}}{{\Gamma \left( \mu  \right){{\left( {\mu \kappa  + m} \right)}^m}{{\bar r}^{2\mu }}}}\exp \left( { - \frac{{\mu (1 + \kappa ){r^2}}}{{{{\bar r}^2}}}} \right){}_1{F_1}\left( {m;\mu ;\frac{{{\mu ^2}\kappa (1 + \kappa )}}{{\mu \kappa  + m}}\frac{{{r^2}}}{{{{\bar r}^2}}}} \right) 
\end{equation}
	 
\noindent where $\Gamma(\bullet)$ is the gamma function and $_{1}F_{1}(\bullet;\bullet;\bullet)$ is the confluent hypergeometric function \cite{12}. In this model, $m$ is allowed to take any value in the range $m \geq 0$ where $m = 0$ corresponds to complete shadowing of the resultant dominant component and $m \rightarrow \infty$ corresponds to no shadowing of the resultant dominant component. Of course when $m = \infty$, the PDF given in (2) becomes equivalent to the $\kappa-\mu$ PDF given in \cite{3}, whereas when $m = 0$ and hence $\kappa = 0$, the PDF given in (2) reduces to the Nakagami PDF \cite{4}.

\section{Level Crossing Rate}
The level crossing rate of a fading signal envelope, $N_{R}(r)$ is defined as the expected number of times that the envelope crosses a given signal level in a positive (or negative) direction per second and is given by \cite{13}

\begin{equation}
N_{R}(r) = \int^{\infty}_{0} \dot{r}f_{R, \dot{R}}(r,\dot{r})d\dot{r}
\end{equation}

\noindent where $\dot{r}$ is the time derivative of $r$ and $f_{R, \dot{R}}(r,\dot{r})$ is the joint probability density of $R$ and $\dot{R}$. If we initially hold the shadowing fluctuation constant, the variation of the signal envelope would follow a $\kappa-\mu$ distribution \cite{3}. In this instance, from (1), it is easy to see that the $\kappa-\mu$ signal power can be obtained as the sum of $\mu$ squared Rice variates. This is a simple but important observation as it allows us to show that the PDF of the time derivative of $R$, denoted as $\dot{R}$, is zero-mean Gaussian distributed. Letting $Z$ represent a Rice distributed random variable, it follows that

\begin{equation}
R^{2} = \sum^{\mu}_{i=1} Z^{2}_{i}.
\end{equation}

\noindent Differentiating both sides of (4) with respect to time we find that

\begin{equation}
\dot{R} = \frac{\sum^{\mu}_{i=1} Z_{i}\dot{Z}_{i}}{R}.
\end{equation}

\noindent Knowing that for the Rice case, $\dot{Z}$ is a zero mean Gaussian distributed random variable with variance $\dot{\sigma}^2_{Z} = 2 \pi^{2}f^{2}_{m}\sigma^{2}$ \cite[eq. (2.104)]{14}, where $f_{m}$ is the maximum Doppler frequency, it is straightforward to show that in fact $\dot{\sigma}^2_{R} \equiv \dot{\sigma}^2_{Z}$. From [3], $\sigma^{2} = \bar{r}^{2} / 2 \mu (1 + \kappa)$ and therefore

\begin{equation}
\dot{\sigma}^{2}_{R} = \frac{\pi^{2}f^{2}_{m}\bar{r}^{2}}{\mu (1 + \kappa)}.	
\end{equation}

\noindent Most importantly though, from (5), we can see that $\dot{R}$ is obtained as a linear transformation of $\dot{Z}$ and thus it can be deduced in a similar fashion to \cite{13} that the PDF of the rate of change of the envelope $\dot{R}$ is uncorrelated with $R$ and thus $f_{R, \dot{R}}(r,\dot{r}) = f_{R}(r)\times f_{\dot{R}}(\dot{r})$.

Now considering the shadowed fluctuation of the dominant component separately, which in this model is assumed to follow a Nakagami-$m$ distribution. Using the model given in \cite{15} it has already been shown that the slope is zero mean Gaussian distributed and its PDF independent of the envelope and thus $f_{R, \dot{R}}(r,\dot{r}) = f_{R}(r)\times f_{\dot{R}}(\dot{r})$. Knowing that the Nakagami-$m$ distribution appears as a special case of the $\kappa-\mu$ distribution, the variance of the slope can also be obtained by letting $\kappa$ = 0 in (6), $\mu = m$ and interchanging $\bar{r}^2$ with $\Omega$ such that $\dot{\sigma}^{2}_{R} = {\pi^{2}f^{2}_{m}\Omega}/{m}$. As above, to remove the dependency of the formulations on the mean power of the dominant component, we substitute $\Omega = \kappa\bar{r}^{2}/(1+\kappa)$ which gives

\begin{equation}
\dot{\sigma}^{2}_{R} = \frac{\pi^{2}f^{2}_{m}\kappa\bar{r}^{2}}{m(1+\kappa)}.	
\end{equation}

Having demonstrated that for both the multipath and shadowing, the variation of the fading components are independent of the PDF of the time derivative of the envelope, it now becomes possible to rewrite (3) as

\begin{equation}
N_{R}(r) = f_{R}(r)\int^{\infty}_{0} \dot{r}f_{\dot{R}}(\dot{r})d\dot{r}
\end{equation}

\noindent where $f_{R}(r)$ is the $\kappa-\mu$ shadowed PDF given in (2) and $f_{\dot{R}}(\dot{r})$ is the PDF of the rate of change of the envelope $\dot{R}$. Following the approach taken in \cite{7} it seems reasonable to assume that the PDF of $\dot{R}$ is the result of two correlated zero-mean Gaussian random processes. Letting $\dot{R} = \dot{A} + \dot{B}$ where $\dot{A}$ is the rate of change of the envelope due to the multipath component and $\dot{B}$ is the rate of change of the envelope due to the shadowed dominant component, the joint density of $\dot{A}$ and $\dot{B}$ is given by \cite{16}

\begin{equation}
\begin{split}
f_{\dot{A},\dot{B}}(\dot{a},\dot{b}) = \frac{1}{2 \pi \dot{\sigma}_{A}\dot{\sigma}_{B}\sqrt{1-\rho^{2}}}
\times
\exp \left [ -\frac{1}{2(1-\rho^{2})}
\left(\frac{\dot{a}^2}{\dot{\sigma}^2_{A}} - \frac{2 \rho \dot{a} \dot{b}}{\dot{\sigma}_{A} \dot{\sigma}_{B}} +\frac{\dot{b}^2}{\dot{\sigma}^2_{B}}\right)\right ] \hspace{10 mm} {|\rho| < 1}
\end{split}
\end{equation}

\noindent where $\dot{\sigma}_{A}$ and $\dot{\sigma}_{B}$ are the variances of the two random variables $\dot{A}$ and $\dot{B}$, and $\rho$ is the correlation between them. Substituting $\dot{A} = \dot{R} - \dot{B}$ into (9), the integral $f_{\dot{R}}(\dot{r}) = \int^{\infty}_{-\infty} f_{\dot{A}, \dot{B}}(\dot{r} - \dot{b}, \dot{b}) d\dot{b}$ can be evaluated as \cite{7}

\begin{equation}
\begin{split}
f_{\dot{R}}(\dot{r}) = \frac{1}{[ 2 \pi (1 - \rho ^ {2}) (\dot{\sigma}^2_{A} + 2 \rho \dot{\sigma}_{A} \dot{\sigma}_{B} + \dot{\sigma}^2_{B})]^{1/2}}
\times
\exp \left[ - \frac{\dot{r}^2}{2 (1-\rho^{2})\dot{\sigma}^2_{A}} \left( \frac{\dot{\sigma}^2_{A}(1-\rho^{2})+4 \rho \dot{\sigma}_{A} \dot{\sigma}_{B} }{\dot{\sigma}^2_{A} + 2 \rho \dot{\sigma}_{A} \dot{\sigma}_{B} + \dot{\sigma}^2_{B}} \right ) \right ]
\end{split}
\end{equation}

\noindent and therefore

\begin{equation}
\begin{split}
\int_0^\infty  {\dot r} {f_{\dot R}}(\dot r)d\dot r = \frac{{\sqrt {(1 - {\rho ^2})\left( {\dot \sigma _A^2 + 2\rho {{\dot \sigma }_A}{{\dot \sigma }_B} + \dot \sigma _B^2} \right)} \dot \sigma _A}}{{\sqrt {2\pi } \left( {\dot \sigma _A(1 - {\rho ^2}) + 4\rho {{\dot \sigma }_B}} \right)}}.
\end{split}
\end{equation}

\noindent Substituting (2), (6), (7) and (11)\footnote{(6) is used in place of $\dot{\sigma} _A^2$ while (7) is used in place of $\dot{\sigma} _B^2$} into (8) and performing the necessary mathematical operations, we obtain the LCR of the $\kappa-\mu$ shadowed fading envelope (normalized to the maximum Doppler frequency, $f_{m}$) as

\begin{equation}
\begin{split}
\frac{{{N_R}(r)}}{{{f_m}}} = \frac{{\sqrt {2\pi \left( {1 - {\rho ^2}} \right)} {\mu ^{\mu  - \frac{1}{2}}}{m^m}{{\left( {1 + \kappa } \right)}^{\mu  - \frac{1}{2}}}{{\left( {m + \mu \kappa  + 2\rho \sqrt {\mu \kappa m} } \right)}^{\frac{1}{2}}}}}{{\Gamma \left( \mu  \right){{\left( {\mu \kappa  + m} \right)}^m}\left( {\sqrt m \left( {1 - {\rho ^2}} \right) + 4\rho \sqrt {\mu \kappa } } \right)}}{\left( {\frac{r}{{\bar r}}} \right)^{2\mu  - 1}}
\\ 
\times \exp \left( { - \frac{{\mu \left( {1 + \kappa } \right){r^2}}}{{{{\bar r}^2}}}} \right){}_1{F_1}\left( {m;\mu ;\frac{{{\mu ^2}\kappa \left( {1 + \kappa } \right)}}{{\mu \kappa  + m}}{{\left( {\frac{r}{{\bar r}}} \right)}^2}} \right)
\end{split}
\end{equation}

\noindent For the case when the slopes of the multipath and shadowed components of the received signal are uncorrelated (i.e. $\rho = 0$), (12) can be further reduced to

\begin{equation}
\begin{split}
\frac{{{N_R}(r)}}{{{f_m}}} = \frac{{\sqrt {2\pi } {\mu ^{\mu  - \frac{1}{2}}}{m^{m  - \frac{1}{2}}}{{\left( {1 + \kappa } \right)}^{\mu  - \frac{1}{2}}}{{\left( {m + \mu \kappa } \right)}^{\frac{1}{2}}}}}{{\Gamma \left( \mu  \right){{\left( {\mu \kappa  + m} \right)}^m}}}{\left( {\frac{r}{{\bar r}}} \right)^{2\mu  - 1}}
\\ 
\times \exp \left( { - \frac{{\mu \left( {1 + \kappa } \right){r^2}}}{{{{\bar r}^2}}}} \right){}_1{F_1}\left( {m;\mu ;\frac{{{\mu ^2}\kappa \left( {1 + \kappa } \right)}}{{\mu \kappa  + m}}{{\left( {\frac{r}{{\bar r}}} \right)}^2}} \right)
\end{split}
\end{equation}

\noindent Fig. 1 shows the shape of the normalized $\kappa-\mu$ shadowed LCR given in (12) for increasing values of $m$ (continuous lines) and decreasing values of $\rho$ (dashed lines). It is quite evident that as the amount of shadowing of the resultant dominant component decreases, i.e. $m$ gets larger, the signal crosses lower levels at lower rates. Furthermore, the impact of increasing correlation between the slope of the shadowed dominant and multipath signals also acts to cause the signal to cross lower levels at lower rates. Fig. 2 shows the normalized LCR of the $\kappa-\mu$ shadowed fading signal for the special cases when it coincides with the normalized LCRs of the Nakagami \cite{15}, Rice \cite{14} and $\kappa-\mu$ \cite{17} fading models i.e. $\kappa = m = 0$ for Nakagami, $\mu = 1$ and $m = \infty$ for Rice and $m = \infty$ for $\kappa-\mu$.  

\section{Average Fade Duration}
The average fade duration (AFD) of a fading signal envelope, $T_{R}(r)$, is defined as the average length of time that the signal spends below the threshold level $R$ and is related to the LCR through the relationship \cite{15}

\begin{equation}
T_{R}(r) = \frac{F_{R}(r)}{N_{R}(r)}.	
\end{equation}

\noindent As we can see, to calculate the AFD, it is necessary to have an expression for the cumulative distribution function, $F_{R}(r)$, of the $\kappa-\mu$ shadowed fading signal. As the CDF of the instantaneous SNR in $\kappa-\mu$ shadowed fading channels has conveniently been derived in [1, eq. (6)], to obtain the CDF of the received signal envelope, the same quadratic transformation used to yield the PDF of the fading signal is employed which gives

\begin{equation}
{F_R}(r) = \frac{{{\mu ^{\mu  - 1}}{m^m}{{\left( {1 + \kappa } \right)}^\mu }}}{{\Gamma \left( \mu  \right){{\left( {\mu \kappa  + m} \right)}^m}}}{\left( {\frac{r}{{\bar r}}} \right)^{2\mu }}{\Phi _2}\left( {\mu  - m,m;\mu  + 1; - \frac{{\mu (1 + \kappa ){r^2}}}{{{{\bar r}^2}}}, - \frac{{\mu (1 + \kappa )m}}{{\left( {\mu \kappa  + m} \right)}}\frac{{{r^2}}}{{{{\bar r}^2}}}} \right)
\end{equation}

\noindent where $\Phi_{2}(\bullet,\bullet;\bullet;\bullet,\bullet)$ is the bivariate confluent hypergeometric function. Now, the normalized AFD of a $\kappa-\mu$ shadowed fading signal can be straightforwardly obtained by substituting (12) and (15) into (14) which gives

\begin{equation}
{T_R}(r){f_m} = \frac{{\frac{{\left( {\sqrt m \left( {1 - {\rho ^2}} \right) + 4\rho \sqrt {\mu \kappa } } \right){{\left( {1 + \kappa } \right)}^{\frac{1}{2}}}}}{{\sqrt {2\pi \left( {1 - {\rho ^2}} \right)} {\mu ^{\frac{1}{2}}}{{\left( {m + \mu \kappa  + 2\rho \sqrt {\mu \kappa m} } \right)}^{\frac{1}{2}}}}}\frac{r}{{\bar r}}{\Phi _2}\left( {\mu  - m,m;\mu  + 1; - \frac{{\mu (1 + \kappa ){r^2}}}{{{{\bar r}^2}}}, - \frac{{\mu (1 + \kappa )m}}{{\left( {\mu \kappa  + m} \right)}}\frac{{{r^2}}}{{{{\bar r}^2}}}} \right)}}{{\exp \left( { - \frac{{\mu (1 + \kappa ){r^2}}}{{{{\bar r}^2}}}} \right){}_1{F_1}\left( {m;\mu ;\frac{{{\mu ^2}\kappa (1 + \kappa )}}{{\mu \kappa  + m}}\frac{{{r^2}}}{{{{\bar r}^2}}}} \right)}}
\end{equation}

\noindent Again for the case when the slopes of the multipath and shadowed components of the received signal are uncorrelated (i.e. $\rho = 0$), (16) can be further reduced to

\begin{equation}
{T_R}(r){f_m} = \frac{{\frac{{\sqrt m {{\left( {1 + \kappa } \right)}^{\frac{1}{2}}}}}{{\sqrt {2\pi } {\mu ^{\frac{1}{2}}}{{\left( {m + \mu \kappa } \right)}^{\frac{1}{2}}}}}\frac{r}{{\bar r}}{\Phi _2}\left( {\mu  - m,m;\mu  + 1; - \frac{{\mu (1 + \kappa ){r^2}}}{{{{\bar r}^2}}}, - \frac{{\mu (1 + \kappa )m}}{{\left( {\mu \kappa  + m} \right)}}\frac{{{r^2}}}{{{{\bar r}^2}}}} \right)}}{{\exp \left( { - \frac{{\mu (1 + \kappa ){r^2}}}{{{{\bar r}^2}}}} \right){}_1{F_1}\left( {m;\mu ;\frac{{{\mu ^2}\kappa (1 + \kappa )}}{{\mu \kappa  + m}}\frac{{{r^2}}}{{{{\bar r}^2}}}} \right)}}
\end{equation}

Fig. 3 shows the shape of the $\kappa-\mu$ shadowed AFD for increasing values of $m$ (continuous lines) and decreasing values of $\rho$ (dashed lines). As the shadowing of the dominant component decreases, the fading envelope spends more time at lower threshold levels. This is in direct contrast to the correlation between the time derivative of the multipath and shadowed dominant components. In this instance, as the correlation $increases$, the signal can spend more time at lower threshold levels.

\section{A Comparison with Measured Shadowed Fading Channels}
To illustrate the utility of the new equations for modeling shadowed fading channels, they were compared with data obtained from two different sets of field measurements which considered channels which are known to susceptible to shadowed fading. The first set of measurements considered cellular device-to-device communications channels operating at 868 MHz in an outdoor urban environment. Full details of the experimental setup are available in \cite{2, 18}. This particular scenario considered two persons spaced 10 m apart in an open space between three buildings in a built up residential area in the suburbs of Belfast in the United Kingdom. Both persons were initially stationary, in direct LOS and had the hypothetical user equipment (UE) positioned at their heads. They were then instructed to move around randomly within a circle of radius 1 m from their starting points while imitating a voice call. The first person's UE was configured to transmit data packets at a power level of 0 dBm with a period of 70 ms for 80 s while the second person's UE recorded the received signal strength indicator (RSSI) of the received packets. The second set of field measurements considered on-body communications channels operating at 2.45 GHz within a highly reverberant environment. Full details of the experimental setup are available in \cite{19}. The on-body link spanned the left-waist to right-knee positions and the measurements were made when the person performed walking on the spot movements. In this instance, the complex $S_{21}$ was sampled with a period of 5 ms for an interval of 30 s.

Fig. 4 shows the empirical level crossing rates for both channels compared to the new equation given in (12). All parameter estimates for the $\kappa-\mu$ shadowed fading model were obtained using the \texttt{lsqnonlin} function available in the Optimization toolbox of MATLAB along with the PDF given in (2). It should be noted that both sets of data were normalized to their respective $rms$ signal levels prior to parameter estimation. Using these parameter estimates, the maximum Doppler frequency and correlation were then obtained by minimizing the sum of the squared error between the empirical and theoretical LCR plots. As we can quite clearly see, the normalized LCR of the $\kappa-\mu$ shadowed fading model provides an excellent fit to the on-body data and a very good approximation of the device-to-device channel. To allow the reader to reproduce these plots, parameter estimates for both measurement scenarios are given in Table I. In both applications, it is quite clear that the fading channel is subject to heavy shadowing with $m \leq 0.55$. Furthermore, for the device-to-device fading channel, it is apparent that the correlation between the slope of the shadowed dominant and multipath components is non-zero. For completeness, Fig. 5 shows both the empirical and theoretical AFD of both types of fading channel. Again, the theoretical AFD provides an excellent representation of the measured data for the on-body fading channel and a good fit for the device-to-device channel for signal levels above -10 dB threshold level.

\section{Conclusion}
Closed-form expressions for the LCR and AFD of the recently proposed $\kappa-\mu$ shadowed fading model have been presented. These new, very general equations will find use in a wide variety of existing and emerging communications applications in which the received signal is subject to shadowed fading such as device-to-device communications, body centric communications and land mobile satellite communications. The analytical expressions have been validated through reduction to known special cases. It was found that decreasing shadowing of the resultant dominant component reduces crossings at low signal levels but at the same time may increase fade durations at these levels. Finally the utility of the new formulations has been proven through comparison with empirical data obtained for cellular device-to-device and body centric fading channels. It has been shown that the second-order statistics of the shadowed fading in both types of channel can be adequately described by the equations proposed here.

\section*{Acknowledgment}

The author is extremely grateful to the reviewers of this manuscript and their invaluable comments which have helped to significantly improve the contribution of the work. This work was supported by the U.K. Royal Academy of Engineering, the Engineering and Physical Sciences Research Council under Grant References EP/H044191/1 and EP/L026074/1 and also by the Leverhulme Trust, UK through PLP-2011-061.

\ifCLASSOPTIONcaptionsoff
  \newpage
\fi



\clearpage
\bibliographystyle{IEEEtran}
\bibliography{ref}

\begin{thebibliography}{10}
\providecommand{\url}[1]{#1}
\csname url@samestyle\endcsname
\providecommand{\newblock}{\relax}
\providecommand{\bibinfo}[2]{#2}
\providecommand{\BIBentrySTDinterwordspacing}{\spaceskip=0pt\relax}
\providecommand{\BIBentryALTinterwordstretchfactor}{4}
\providecommand{\BIBentryALTinterwordspacing}{\spaceskip=\fontdimen2\font plus
\BIBentryALTinterwordstretchfactor\fontdimen3\font minus
  \fontdimen4\font\relax}
\providecommand{\BIBforeignlanguage}[2]{{%
\expandafter\ifx\csname l@#1\endcsname\relax
\typeout{** WARNING: IEEEtran.bst: No hyphenation pattern has been}%
\typeout{** loaded for the language `#1'. Using the pattern for}%
\typeout{** the default language instead.}%
\else
\language=\csname l@#1\endcsname
\fi
#2}}
\providecommand{\BIBdecl}{\relax}
\BIBdecl

\bibitem{1}
J.~F. Paris, ``Statistical characterization of $\kappa-\mu$ shadowed fading,''
  \emph{IEEE Transactions on Vehicular Technology}, vol.~63, no.~2, pp.
  518--526, Feb 2014.

\bibitem{2}
S.~L. Cotton, ``Human body shadowing in cellular device-to-device
  communications: channel modeling using the shadowed $\kappa-\mu$ fading
  model,'' \emph{IEEE Journal on Selected Areas in Communications}, vol.~33,
  no.~1, pp. 111--119, Jan 2015.

\bibitem{3}
M.~D. Yacoub, ``The $\kappa-\mu$ distribution and the $\eta-\mu$
  distribution,'' \emph{IEEE Antennas and Propagation Magazine}, vol.~49,
  no.~1, pp. 68--81, Feb 2007.

\bibitem{4}
M.~Nakagami, \emph{The $m$-distribution: A general formula of intensity
  distribution of rapid fading, in Statistical Methods in Radio Wave
  Propagation}.\hskip 1em plus 0.5em minus 0.4em\relax New York: Pergamon,
  1960.

\bibitem{5}
A.~Abdi, W.~C. Lau, M.-S. Alouini, and M.~Kaveh, ``A new simple model for land
  mobile satellite channels: first- and second-order statistics,'' \emph{IEEE
  Transactions on Wireless Communications}, vol.~2, no.~3, pp. 519--528, May
  2003.

\bibitem{6}
J.~C. S.~S. Filho and M.~D. Yacoub, ``Nakagami-\textit{m} approximation to the
  sum of m non-identical independent {Nakagami-\textit{m}} variates,''
  \emph{Electronics Letters}, vol.~40, no.~15, pp. 951--952, July 2004.

\bibitem{7}
C.~Loo, ``A statistical model for a land mobile satellite link,'' \emph{IEEE
  Transactions on Vehicular Technology}, vol.~34, no.~3, pp. 122--127, Aug
  1985.

\bibitem{8}
P.~C. Sofotasios and S.~Freear, ``On the $\kappa-\mu$/gamma composite
  distribution: A generalized multipath/shadowing fading model,'' in \emph{2011
  SBMO/IEEE MTT-S International Microwave Optoelectronics Conference}, Oct
  2011, pp. 390--394.

\bibitem{9}
S.~R. Panic, D.~M. Stefanovic, I.~M. Petrovic, M.~C. Stefanovic, J.~A.
  Anastasov, and D.~S. Krstic, ``Second-order statistics of selection
  macro-diversity system operating over gamma shadowed - fading channels,''
  \emph{Eurasip Journal on Wireless Communications and Networking}, p. 155,
  2011.

\bibitem{10}
N.~Youssef, T.~Munakata, and M.~Takeda, ``Fade statistics in nakagami fading
  environments,'' in \emph{IEEE 4th International Symposium on Spread Spectrum
  Techniques and Applications Proceedings}, vol.~3, Sep 1996, pp. 1244--1247.

\bibitem{11}
S.~L. Cotton, ``A statistical model for shadowed body-centric communications
  channels: Theory and validation,'' \emph{IEEE Transactions on Antennas and
  Propagation}, vol.~62, no.~3, pp. 1416--1424, March 2014.

\bibitem{12}
J.~Abad and J.~Sesma, ``Computation of the regular confluent hypergeometric
  function,'' \emph{Mathematica Journal}, vol.~5, no.~4, pp. 74--76, 1995.

\bibitem{13}
S.~O. Rice, ``Statistical properties of a sine wave plus random noise,''
  \emph{Bell Syst. Tech. Journal}, vol.~27, pp. 109--157, 1948.

\bibitem{14}
G.~L. Stuber, \emph{Principles of Mobile Communication, 3rd Edition}.\hskip 1em
  plus 0.5em minus 0.4em\relax Springer, 2011.

\bibitem{15}
M.~D. Yacoub, J.~E.~V. Bautista, and L.~Guerra~de Rezende~Guedes, ``On higher
  order statistics of the nakagami-\textit{m} distribution,'' \emph{IEEE
  Transactions on Vehicular Technology}, vol.~48, no.~3, pp. 790--794, May
  1999.

\bibitem{16}
A.~Papoulis, \emph{Probability, random variables, and stochastic processes, 3rd
  Edition}.\hskip 1em plus 0.5em minus 0.4em\relax McGraw-Hill, 1991.

\bibitem{17}
S.~L. Cotton and W.~G. Scanlon, ``Higher-order statistics for $\kappa-\mu$
  distribution,'' \emph{Electronics Letters}, vol.~43, no.~22, pp. 1215--1217,
  Oct 2007.

\bibitem{18}
S.~L. Cotton, ``Channel measurements of {device-to-device} communications in an
  urban outdoor environment,'' in \emph{2014 XXXIth URSI General Assembly and
  Scientific Symposium (URSI GASS)}, Aug 2014, pp. 1--4.

\bibitem{19}
------, ``A statistical characterization of {on-body} fading using the shadowed
  $\kappa-\mu$ fading model,'' in \emph{2014 IEEE Antennas and Propagation
  Society International Symposium (APSURSI)}, July 2014, pp. 717--718.

\end{thebibliography}
%



\clearpage

\begin{table}[t]

\renewcommand{\arraystretch}{1.5}
\caption{Estimated Parameters for Measured Shadowed Fading Channels}
\label{Table1}
\centering

\begin{tabular}{ccccccc}

\hline
 \multicolumn{1}{|c|}{Fading Channel} & \multicolumn{1}{c|}{$\hat{\kappa}$} & \multicolumn{1}{c|}{$\hat{\mu}$} & \multicolumn{1}{c|}{$\hat{\bar{r}}$} & \multicolumn{1}{c|}{$\hat{m}$} & \multicolumn{1}{c|}{$\hat{f}_{m}$} & \multicolumn{1}{c|}{$\hat{\rho}$} \\ \hline

 \multicolumn{1}{|c|}{D2D} & \multicolumn{1}{c|}{1.39} & \multicolumn{1}{c|}{1.78} & \multicolumn{1}{c|}{1.14} & \multicolumn{1}{c|}{0.55} & \multicolumn{1}{c|}{2.40} & \multicolumn{1}{c|}{0.29} \\ \hline

 \multicolumn{1}{|c|}{On-Body} & \multicolumn{1}{c|}{0.66} & \multicolumn{1}{c|}{1.39} & \multicolumn{1}{c|}{1.03} & \multicolumn{1}{c|}{0.36} & \multicolumn{1}{c|}{4.68} & \multicolumn{1}{c|}{0.05} \\ \hline

\end{tabular}
\end{table}


\clearpage

\begin{figure}[!t]
\centering
\includegraphics[width=6.8in]{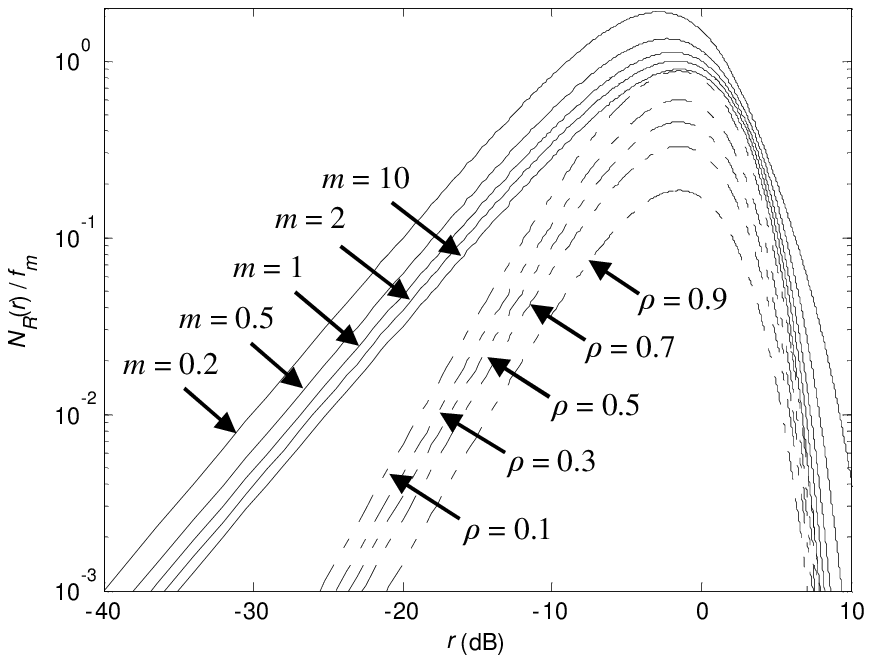} 
\caption{Normalized level crossing rate for the $\kappa-\mu$ shadowed fading model
with decreasing shadowing of the resultant dominant component (continuous lines,
$\rho$ = 0) and with increasing values of the correlation coefficient (dashed
lines, $m$ = 1). It should be noted that for all of the plots, $\kappa$ =
0.5, $\mu$ = 2 and $\bar{r}$ = 1.}
\label{fig_1}
\end{figure}

\begin{figure}[!t]
\centering
\includegraphics[width=6.8in]{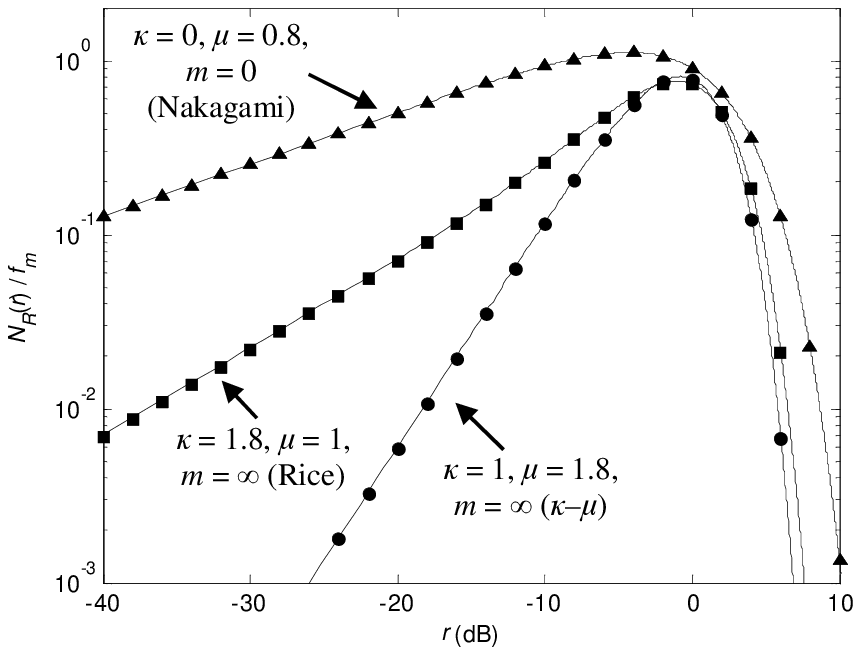} 
\caption{Normalized level crossing rate for the special cases when the
$\kappa-\mu$ shadowed fading model (continuous lines) coincides with the normalized LCR of the
Nakagami \cite{15} (triangle shapes), Rice \cite{14} (square shapes) and
$\kappa-\mu$ \cite{17} (circle shapes) models. It should be noted that for all
of the plots $\bar{r}=1$ and $\rho = 0$ for the $\kappa-\mu$ fading model}.
\label{fig_2}
\end{figure}

\clearpage

\begin{figure}[!t]
\centering
\includegraphics[width=6.8in]{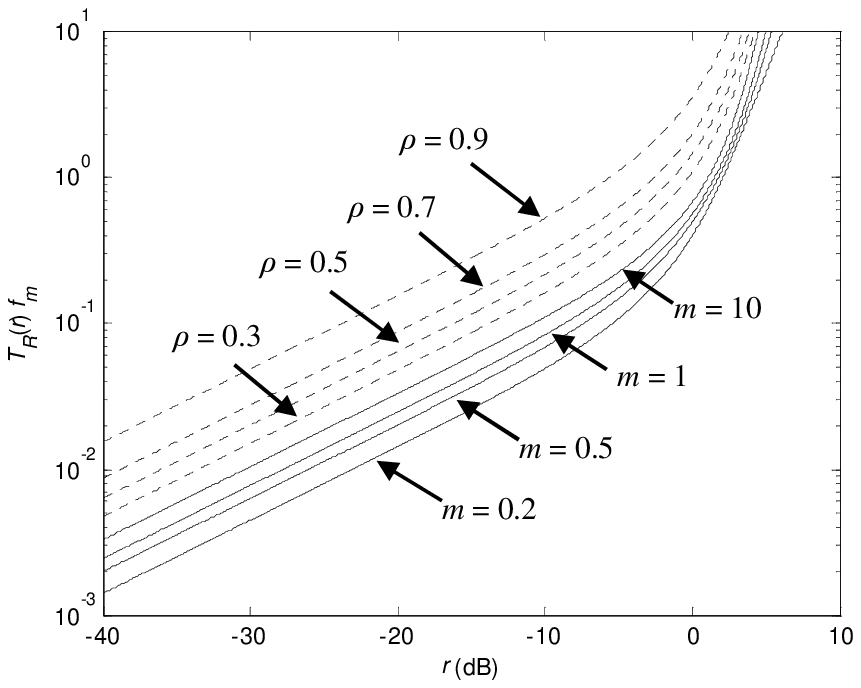} 
\caption{Average fade duration for the $\kappa-\mu$ shadowed fading model
with increasing shadowing of the shadowed dominant component (continuous lines,
$\rho$ = 0) and with increasing values of the correlation coefficient (dashed
lines, $m$ = 1). It should be noted that for all of the plots, $\kappa$ =
0.5, $\mu$ = 2 and $\bar{r}$ = 1.}
\label{fig_3}
\end{figure}

\begin{figure}[!t]
\centering
\includegraphics[width=6.8in]{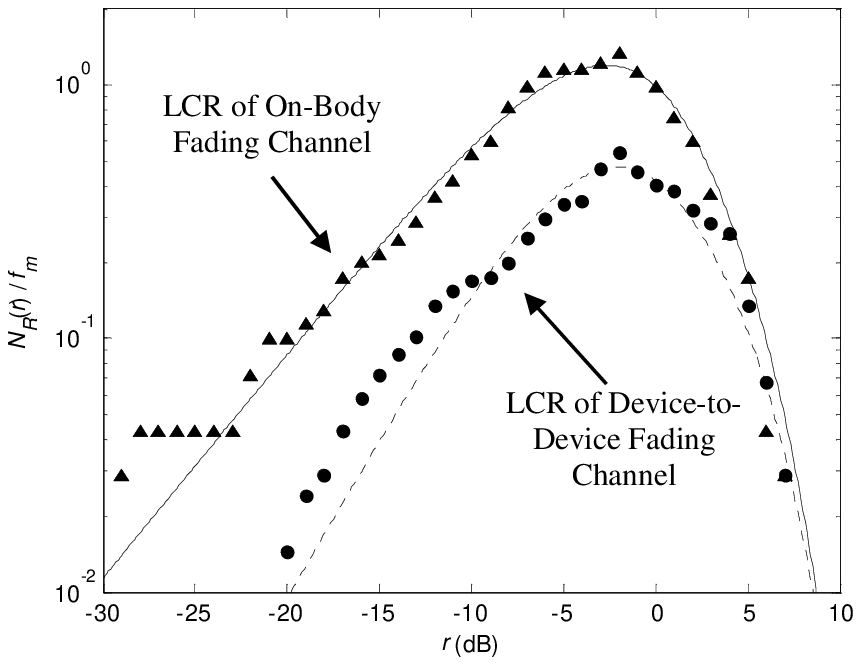} 
\caption{Comparison of the theoretical (continuous lines) and empirical (shapes)
normalized LCRs for the device-to-device and on-body fading channels. All parameter
estimates for the theoretical plots are given in Table~I.}
\label{fig_4}
\end{figure}

\clearpage

\begin{figure}[!t]
\centering
\includegraphics[width=6.8in]{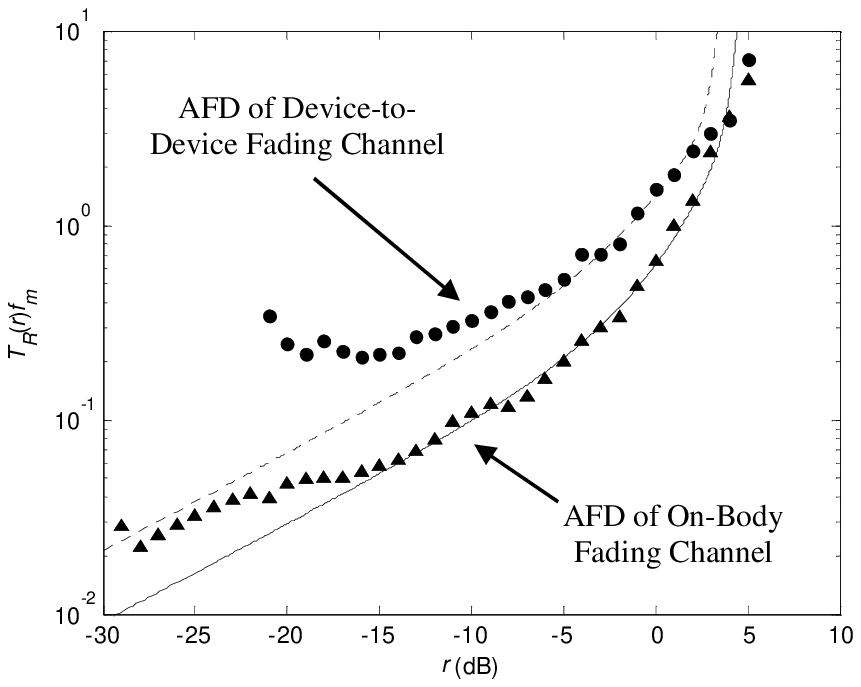} 
\caption{Comparison of the theoretical (continuous lines) and empirical (shapes)
normalized AFDs for the device-to-device and on-body fading channels. All parameter
estimates for the theoretical plots are given in Table~I.}
\label{fig_5}
\end{figure}

\end{document}